\documentclass[11pt]{svproc}
\usepackage{amsmath}
\usepackage{graphicx,subfigure}
\usepackage{amsfonts}
\usepackage{url}
\usepackage{geometry}
\usepackage{booktabs}
\usepackage{geometry}
\usepackage{url}
\usepackage{setspace}
\usepackage[yyyymmdd]{datetime}
\renewcommand\refname{Reference}

\begin{document}

\mainmatter              % start of a contribution
\title{Melody Classification based on Performance Event Vector and BRNN}
\titlerunning{Melody Classification}  % abbreviated title (for running head)
\author{Jinyue Guo\inst{1}\inst{2} \and Aozhi Liu\inst{1} \and Jing Xiao\inst{1}}
\authorrunning{Jinyue Guo et al.} % abbreviated author list (for running head)
\tocauthor{Jinyue Guo, Aozhi Liu, and Jing Xiao}
\institute{Ping An Technology, Shenzhen, China
\and
New York University}

\maketitle              % typeset the title of the contribution

\begin{abstract}
We proposed a model for the CSMT2020 data challenge of melody classification. Our model used the Performance Event Vector as the input sequence to build a Bidirectional RNN network for classfication. The model achieved a satisfying performance on the development dataset and Wikifonia dataset. We also discussed the effect of several hyper-parameters, and created multiple prediction outputs for the evaluation dataset.
% We would like to encourage you to list your keywords within
% the abstract section using the \keywords{...} command.
\end{abstract}
\section{Data Preparation}
To form a classification problem, a balanced human-composed dataset is needed. Furthermore, we need to make sure the distribution of the open-source dataset is similar enough with the distribution of the test set, which means monophonic melody samples are the most suitable for the task.
	
Wikifonia\footnote{Wikifonia website: www.wikifonia.org; Download: http://www.synthzone.com/files/Wikifonia/Wikifonia.zip} is an open-source song dataset which contains more than 6,000 samples of human-composed song samples, including the melody, lyrics, and chord progression. We assumed that these melodies have musicological similarities with the real melodies in the evaluation set, e.g. tonality, rhythm, phrasing, repetition, etc. We parse the original MusicXML samples in Wikifonia into melody-only, 16-bar MIDI files to create our dataset used for feature extraction.
\section{Feature Extraction}
There are two main reasons to use a handcrafted feature, rather than the original MIDI file, as the input of our classification model. Firstly, While MIDI files are specifically encoded musical messages, most machine learning models take pure numerical tensors as the input, which means a ”translation” process is needed. Secondly, since the MIDI protocol is a hardware-level transmission protocol, it contains low-level information that does not encode any musical information. Since our task is to classify the melody, we want our feature to focus on meaningful \textbf{music events}.	
\subsection{Piano Roll Vector}
	
The piano roll vector is a simple representation of music that translates music scores into matrices. We chose not to use piano roll vector for two reasons: as we can see in \ref{fig:reps1}, it becomes indistinguishable for a single quarter note and two repetitive eighth notes. Secondly, while the Wikifonia contains rest notes, which is represented as MIDI pitch 0 in piano roll vector, the development set contains no rest at all. This means any classification model can easily classify a sample by just look at whether it contains 0 or not.
	
\subsection{Performance Event Vector}
	
Introduced by \cite{simon2017performance}, the Performance Event Vector is an \textbf{event based} representation of music. It encodes events such as \textit{NOTE\_ON}, \textit{NOTE\_OFF}, \textit{TIME\_SHIFT}, \textit{VELOCITY\_CHANGE} as discrete events and forms a vocabulary with finite length. We can then apply Natural Language Processing methods to perform classification or even generation tasks. 
	
\begin{figure}[htb]
	    \centering
	    \includegraphics[width=4.2in]{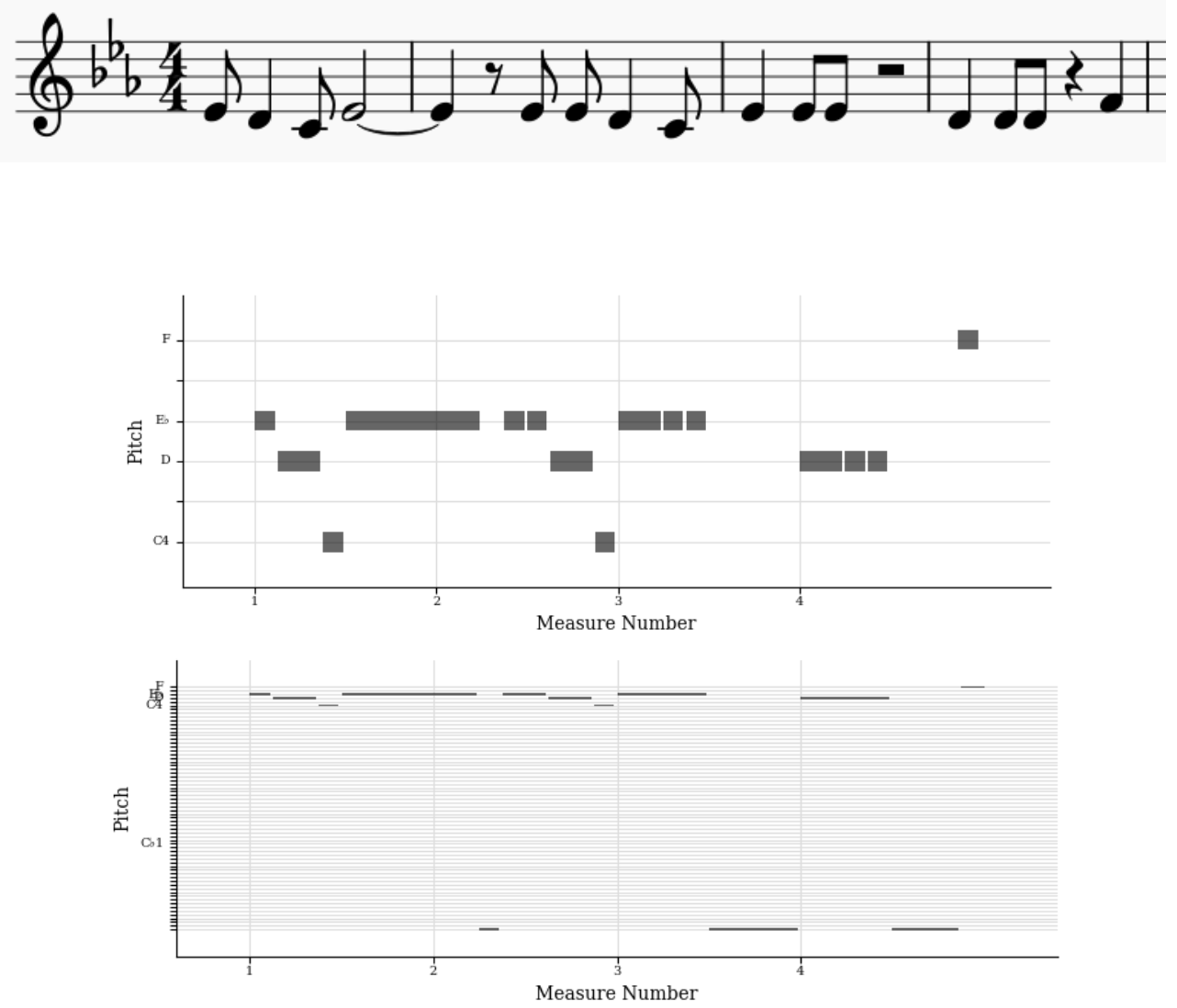}
	    \caption{From top to bottom: a) a sample of score from \textit{Wikifonia}. b) directly translated MIDI file. c) translated by piano roll vector with \textit{col\_fs}=8. Notice all the rests are represented as MIDI note 0, and repeating notes are connected.}
	    \label{fig:reps1}
	\end{figure}

\begin{figure}[htb]
	    \centering
	    \includegraphics[width=5.2in]{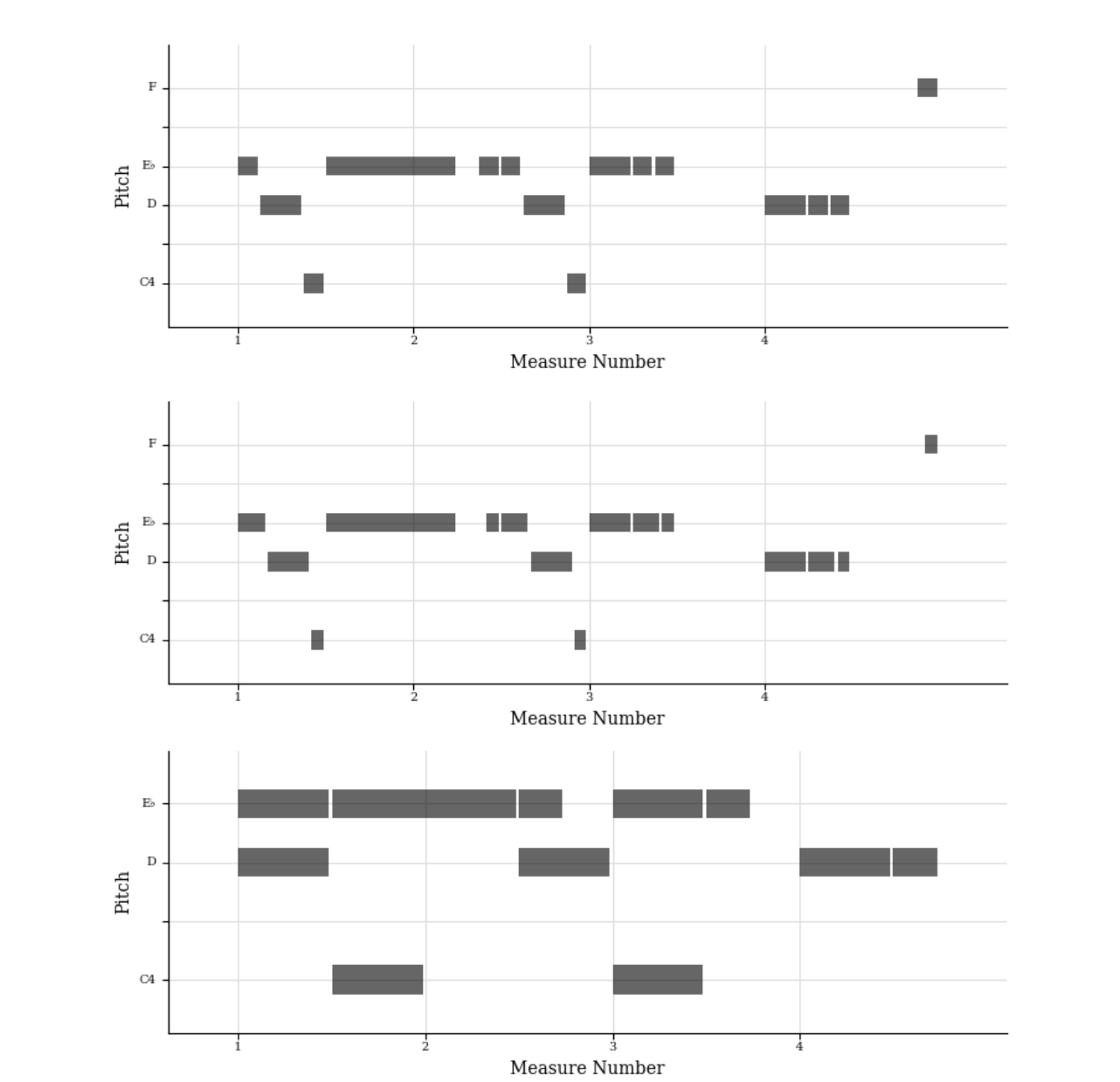}
	    \caption{From top to bottom: a) translated by performance event vector with \textit{steps\_per\_second}=100. b) performance event vector with \textit{steps\_per\_second}=10. c) \textit{steps\_per\_second}=1.}
	    \label{fig:reps2}
	\end{figure}

In our challenge, Since the development dataset does not contain any velocity information, we only encoded three types of events: \textit{NOTE\_ON} and \textit{NOTE\_OFF} for MIDI pitch in range [21, 108], and \textit{TIME\_SHIFT} at different resolutions. Fig. \ref{fig:reps1} and Fig. \ref{fig:reps2} compares the re-generated midi files of different time shift resolutions.
	
We represent the time shift resolution as hyperparameter \textit{steps\_per\_second}. For example, a value of 100 means we divide time-shift events in one second into 100 discrete values, so each event represents a period of integer times of 10ms. A time shift longer than one second will be split into multiple consecutive events.
	
Fig. \ref{fig:reps2} shows how \textit{steps\_per\_second} can affect the translated result. When \textit{steps}=100, the melody can be restored almost perfectly. When \textit{steps}=10, some notes are quantized, and their duration has been changed. When \textit{steps}=1, it could be considered as a short-time pitch histogram with a window size of one second. The effect of this hyperparameter will be further discussed in later chapters.

    \section{Model Structure and Training}
    
    Our model takes a numerical sequence as the input, which is passed to an embedding layer, followed by a recurrent layer. We implemented three variations of the recurrent layer: Bi-directional Long Short-Term Memory (BiLSTM) model\cite{schuster1997bidirectional}, multiplicative Long Short-Term Memory (mLSTM) model\cite{krause2016multiplicative}, and the Recurrent Gate Unit (GRU) model\cite{cho2014GRU}. The last hidden state of two directions are taken as the latent code vector, and passed through a dropout layer. Finally, two dense layers form the single output within range [0, 1] as the final score.
    
	\begin{figure}[htb]
	    \centering
	    \includegraphics[width=5.2in]{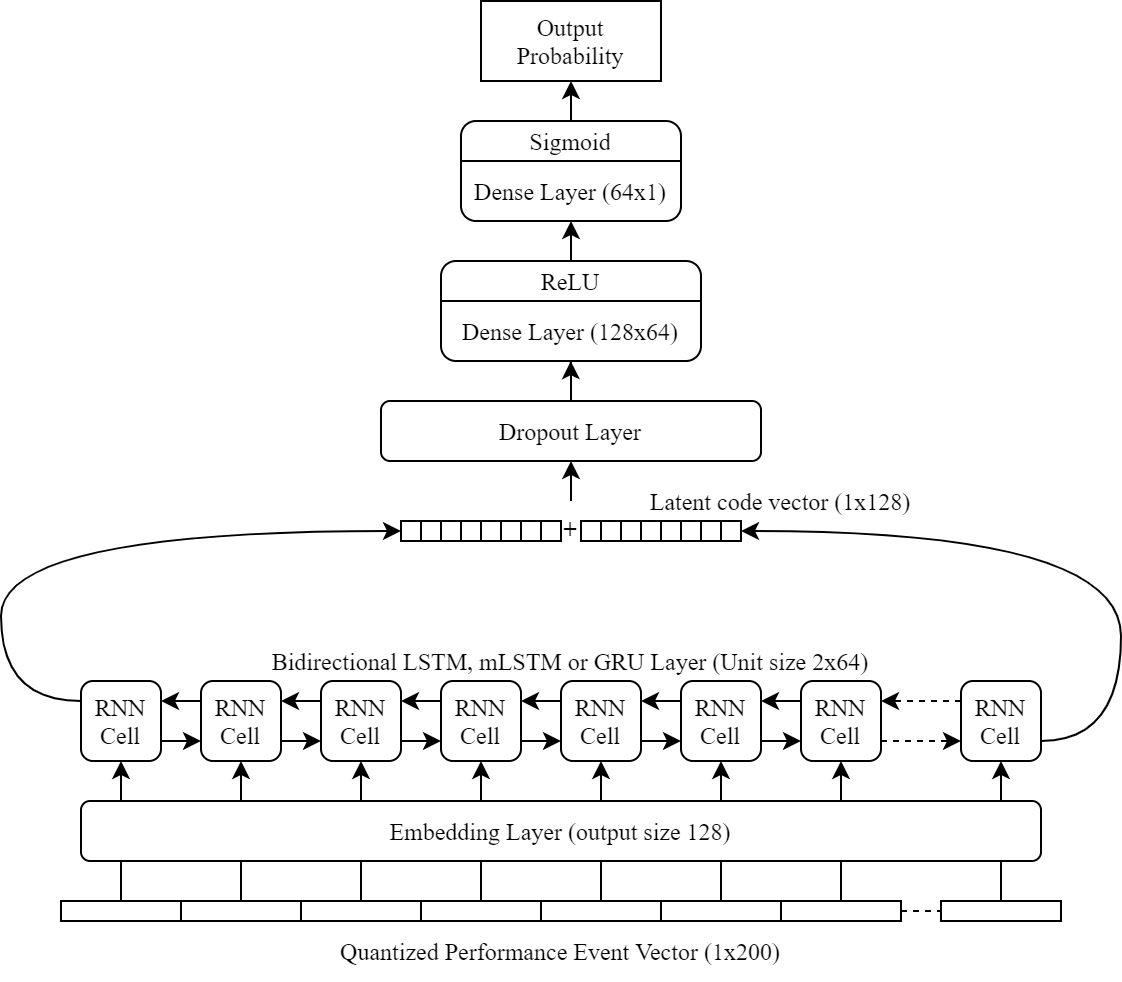}
	    \caption{model structure}
	    \label{fig:model}
	\end{figure}
	
	We trained our model with different performance event hyperparameters, but some parameters are identical. We set our pitch range from MIDI note 21 to 108, number of velocity bins as 0, since both the development data and Wikifonia data do not contain any velocity information. The sequence length is set to 200, a sample has more than 200 events will be truncated, while samples shorter than 200 will be padded with zeros.
	
	We manually chose seven values of \textit{steps\_per\_second} from 100 to 1 and trained models accordingly. These values are 100, 50, 20, 10, 5, 2, and 1. The change of \textit{steps\_per\_second} will affect the number of time-shift events, thus changing the total size of the event vocabulary. The size of vocabulary can be calculated using Eq. \eqref{eq:1}:

	\begin{equation}
	\begin{split}
		vocab\_size & = 2\times(MAX\_PITCH-MIN\_PITCH+1) \\
		            & \qquad  +STEPS\_PER\_SECOND+NUM\_VELOCITY\_BINS+2 
	\end{split}
	\label{eq:1}
	\end{equation}
	
	The first two events in vocabulary are reserved for padding (0) and EOS (1).
	
	The embedding layer's input size is set to the corresponding vocabulary size, and output size of 128. Recurrent units have a unit size of 64, which is doubled if bi-direction is used. The input dropout ratio of recurrent layer is 0.2, but the recurrent dropout ratio is 0. The size of latent code vector is 64 for single direction and 128 for bi-direction. The dropout layer's ratio is 0.2.
	
\begin{figure}[hb]
	    \centering
	    \includegraphics[width=5.2in]{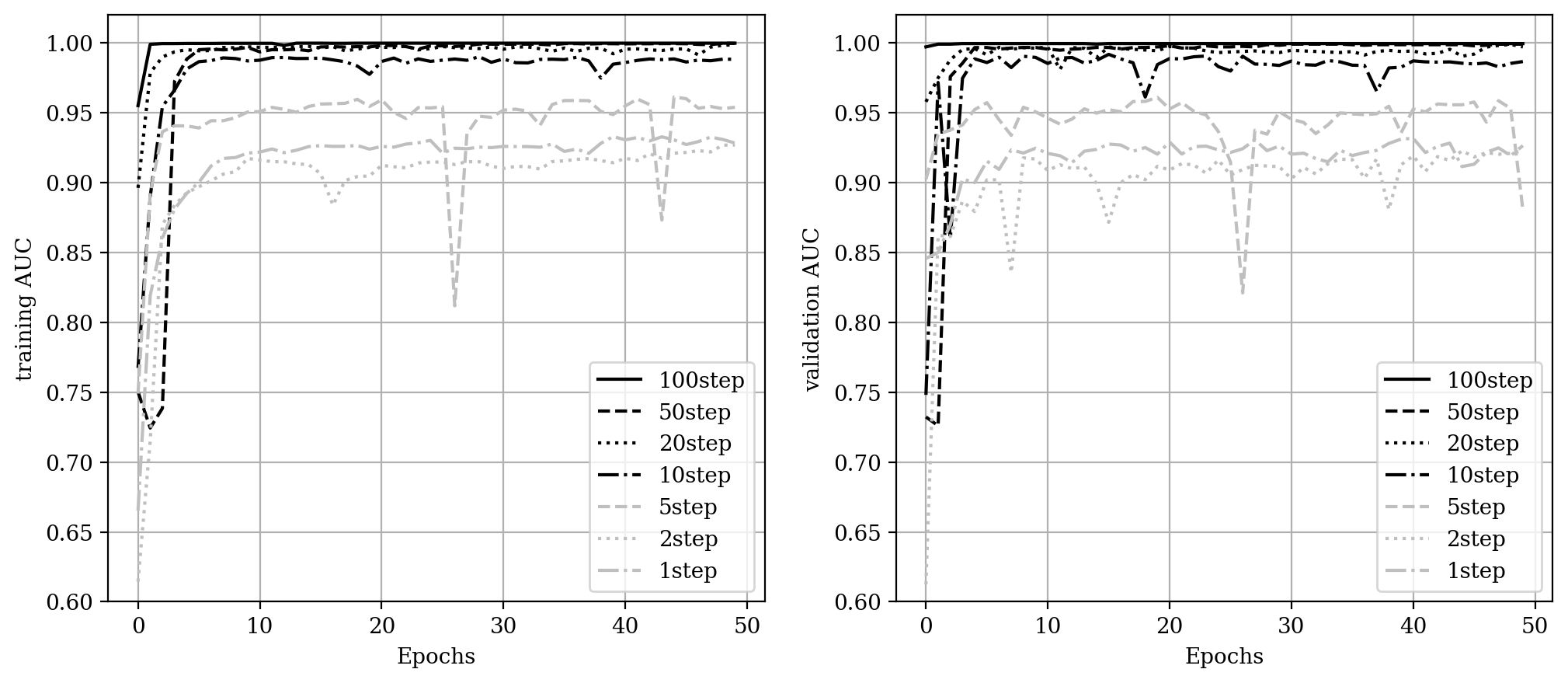}
	    \caption{training and validation AUC}
	    \label{fig:AUC}
	\end{figure}

	The models are trained for 50 epochs with binary cross-entropy loss function, Adam optimizer with a learning rate of 0.0001, and a validation split of 0.2. Fig. \ref{fig:AUC} shows the training and validation AUC of different steps.

    \section{Hyperparameter selection and analysis}
    
    While different recurrent layers give the similar result, different steps settings can hugely affect the result. We can see a 1.0 AUC for 100step model, and the AUC keeps to decline when the step number becomes smaller. When step number comes to 1, the AUC drops to around 0.93. Every model learns quickly, their performance are all stablized after 10 epochs.
    
    However, we need to be more careful when looking at the training and validation result. Even though more steps seems to have a significantly better AUC, we can not simply conclude that they are better at predicting the evaluation dataset. The reason is that we have assumed that the Wikifonia dataset has the same distribution as the real samples in the evaluation data, which might not be true. 
    
    Since our Wikifonia data is generated by MusicXML files, every note has a perfectly quantized start point and duration, while we have no clue how the real samples in the evaluation dataset are generated. If the real samples were recorded from live performance, then the timestamps might not be perfect, which could lead to misclassification if we set a high number of steps. On the other hand, since the low step models will quantize each note strictly, they can ignore the time stamp features and might be able to focus on learning musical features that can distinguish human-composed music and AI-composed music.
    
    Therefore, We included the prediction result for each of the seven step settings in our submission. To prevent overfitting, we used the first checkpoint when the training AUC is stabilized, which is 10 epochs for all step settings. 

\bibliographystyle{gbt-7714-2015-numerical}
\bibliography{mybib}

\end{document}